\documentclass[9pt, journal]{llncs}

\usepackage{graphicx,epsfig,amsmath}

\DeclareMathOperator{\minimize}{minimize}
\DeclareMathOperator{\st}{subject\ to}

\begin{document}

\title{Locally-Optimized Inter-Subject Alignment of Functional Cortical Regions}

\author{Marius C\u{a}t\u{a}lin Iordan\inst{1} \and Armand Joulin\inst{1} \and Diane M. Beck\inst{2} \and Li Fei-Fei\inst{1}}

\institute{Computer Science Department, Stanford University \and Beckman Institute and Department of Psychology, University of Illinois at Urbana-Champaign}

\maketitle

\begin{abstract}
Inter-subject registration of cortical areas is necessary in functional imaging (fMRI) studies for making inferences about equivalent brain function across a population. However, many high-level visual brain areas are defined as peaks of functional contrasts whose cortical position is highly variable. As such, most alignment methods fail to accurately map functional regions of interest (ROIs) across participants. To address this problem, we propose a locally optimized registration method that directly predicts the location of a seed ROI on a separate target cortical sheet by maximizing the functional correlation between their time courses, while simultaneously allowing for non-smooth local deformations in region topology. Our method outperforms the two most commonly used alternatives (anatomical landmark-based AFNI alignment and cortical convexity-based FreeSurfer alignment) in overlap between predicted region and functionally-defined LOC. Furthermore, the maps obtained using our method are more consistent across subjects than both baseline measures. Critically, our method represents an important step forward towards predicting brain regions without explicit localizer scans and deciphering the poorly understood relationship between the location of functional regions, their anatomical extent, and the consistency of computations those regions perform across people.
\end{abstract}

\section{Introduction}
A common and reasonable assumption of modern neuroscience is that virtually all human brain areas, whether functionally or anatomically defined, are shared across the vast majority of the population and a correspondence of processing role exists between such equivalent areas. However, no two brains have the same anatomical shape or folding pattern, and thus finding a precise correspondence between locations in two separate cortical surfaces is a highly non-trivial problem.

Currently, most state-of-the-art cortical prediction and alignment methods define transformations between entire cortical volumes that attempt to preserve anatomical landmarks, cortical curvature, or functional connectivity, and subsequently check whether specific regions of interest (ROIs) are accurately matched between subjects~\cite{Yeo2011,Sabuncu2010,Conroy2013}. However, many high-level visual brain areas are defined as peaks of functional contrasts (e.g. higher activation for objects versus scrambled objects for lateral occipital complex LOC~\cite{GrillSpector1998}) and it is usually difficult to identify clear anatomical landmarks and boundaries for these areas, due to large variability in their cortical position~\cite{Vinberg2008,Amano2009} and functional response~\cite{Arizpe2014} (Fig.~1, left). As a consequence, although they provide a reasonable global matching, previous methods usually fail to accurately map such functional ROIs across participants.

\begin{figure}[t]
\begin{center}
\includegraphics[width=0.45\textwidth]{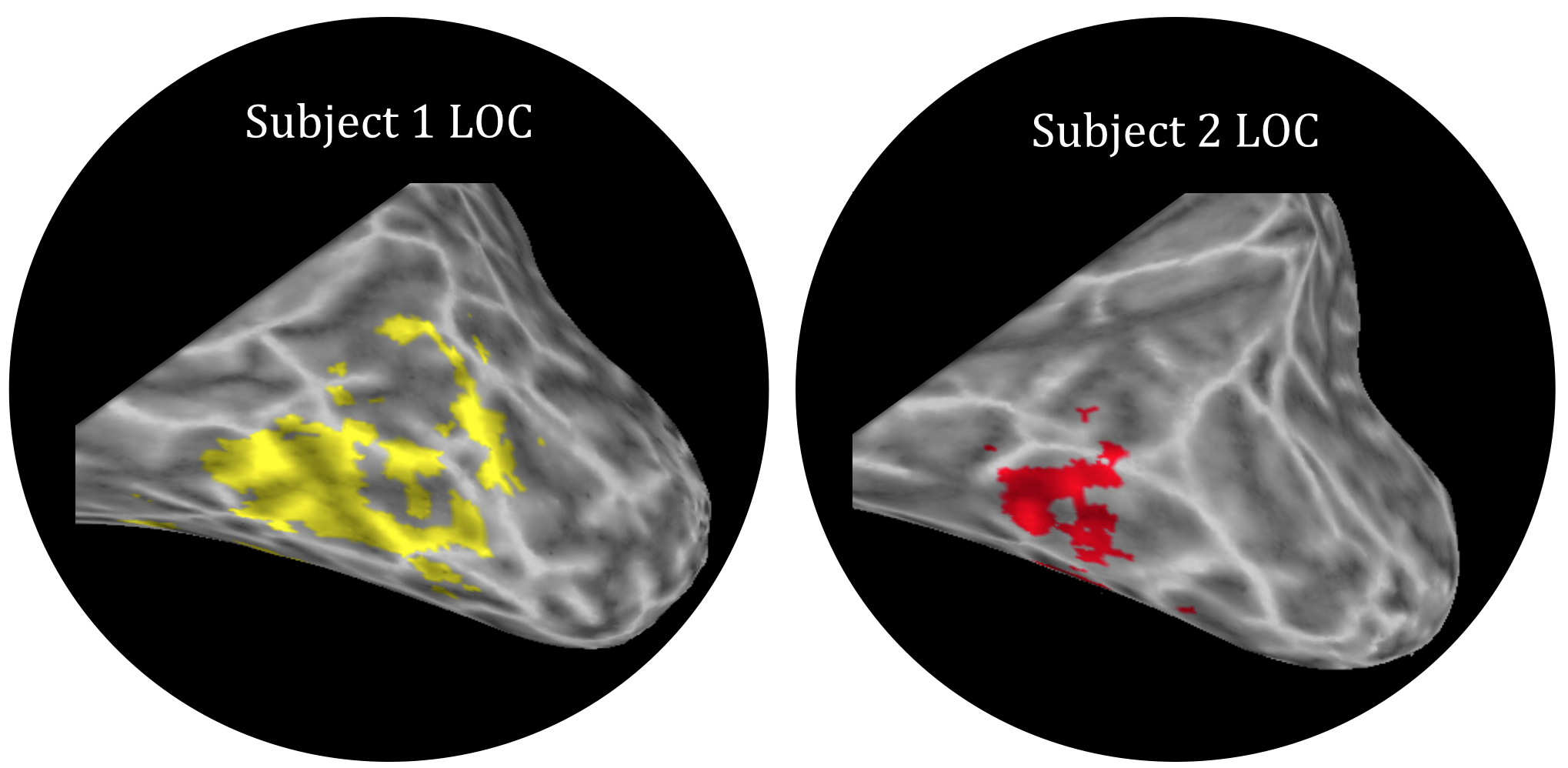}
\hspace{5mm}
\includegraphics[width=0.42\textwidth]{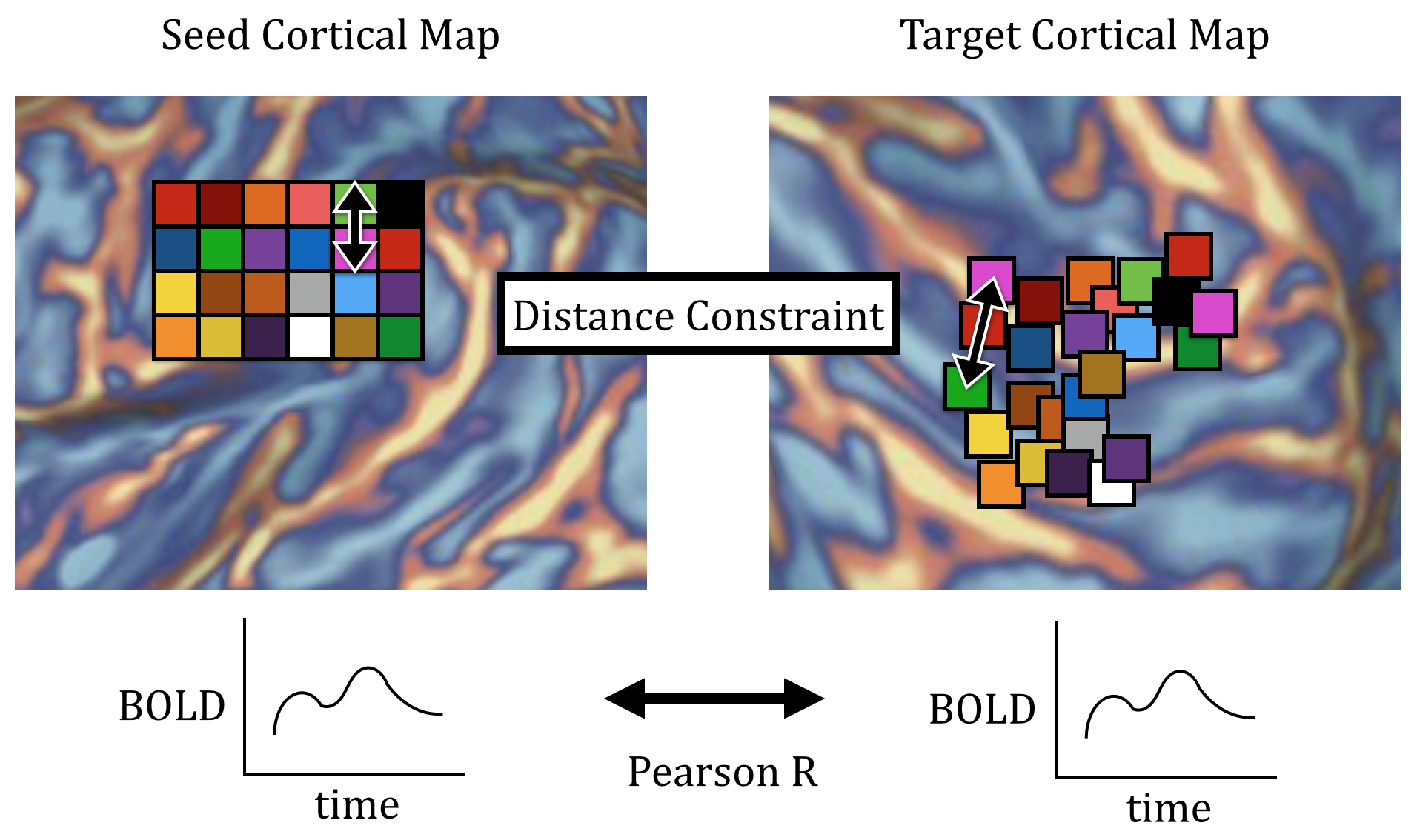}
\caption{\label{stimuli} {\bf (Left) LOC variability.} Location and extent of lateral occipital complex (LOC) is highly variable across subjects, even when using the same localizer experiment, same scanner, and same analysis pipeline.  {\bf (Right) Schematic representation of our proposed method.} Our algorithm tiles the seed region with smaller sub-regions and finds the best functional match for each of them in the target map. The sub-regions are allowed to move independently from one another, provided only that the distance between any two initially adjacent sub-regions does not increase by more than a set threshold.}
\end{center}
\end{figure}

Thus, our goal is to increase the reliability of inter-subject mapping for these cortical functional peaks, as well as for the visual areas they define, using fMRI data. To address this problem, we describe a general-purpose method for predicting the location of functional areas across people that we apply to the problem of localizing object-selective cortex, LOC. Section~2 describes previous work, Section~3 details the algorithm, and the results are shown in Section~4. Finally, Section~5 discusses potential applications.

\section{Related Work}

Our problem can be thought of as a special instance of cortical alignment, where the main goal becomes accurate prediction of a particular region's location, rather than finding a complete correspondence between entire brain volumes. By comparison, virtually all extant alignment methods (\cite{Yeo2011,Sabuncu2010,Conroy2013,Talairach,AFNI,FreeSurfer,Haxby2011,Rustamov2013}) define transformations across full cortical volumes and subsequently check whether specific regions of interest (ROIs) are accurately matched between subjects.

Anatomical alignment relies on large scale correspondences between all human brains, including the reliable presence and the relatively consistent position of primary features such as major sulci and gyri on the cortical surface (e.g. Talairach~\cite{Talairach}, AFNI~\cite{AFNI}). Additionally, given that the main obstacle in aligning the cortical surface between subjects is its folding variability, methods have been proposed that warp gray matter meshes using local curvature properties of the cortex (e.g. FreeSurfer~\cite{FreeSurfer}). These methods, as well as recent extensions~\cite{Yeo2011} also suffer from significant shortcomings in matching functional areas.

A recent method incorporates functional connectivity constraints in the mapping~\cite{Conroy2013} and shows improved ability to align intertwined networks in the brain (i.e. {\em default mode network}). However, many functional areas are not usually a strong part of these networks and thus receive little benefit from this approach.

Finally, another class of alignment methods uses functional correlation constraints. For example, hyperalignment~\cite{Haxby2011,Rustamov2013} and other methods that rely on low-dimensional embeddings of functional responses (e.g.~\cite{Langs2014}) usually offer improvements over commonly used anatomical alignment methods (e.g. Talairach~\cite{Talairach}, AFNI~\cite{AFNI}). Nevertheless, such methods represent a point in the target map as a linear combination of (possibly) all voxels in the other map, and thus are not directly amenable to transferring the location of one contained area across maps without explicit additional knowledge, such as post-hoc labeling. Another promising recent method~\cite{Sabuncu2010} starts with FreeSurfer alignment and maximizes local functional correlation across the cortical surface to {nudge} the vertices of the surface map. This method performs well for early visual areas, but shows limited ability to match functional regions as distance from the occipital pole increases. In contrast to~\cite{Sabuncu2010}, we enforce maximal alignment and prediction specificity to a single region of interest and, furthermore, we allow for locally non-smooth deformations in our mapping, which bypasses the (otherwise ubiquitous in previous work) expectation of using continuous maps between cortical sheets or volumes.

\section{Locally-Optimized Cortical Region Prediction}

Our goal is to predict the location of functionally-defined high-level visual areas between participants. To compute a correspondence between equivalent functional regions, we reasoned that although two cortical surfaces (corresponding to two separate subjects) must express the same necessary computational units that give rise to observed function, these units might not be perfectly equivalent or identically distributed spatially across the two ROIs~\cite{Huth2012}. Thus, a key design principle behind our method is to allow a small degree of non-smoothness in the local deformations afforded by the mapping between the two cortical surfaces.

Our method was inspired by a computer vision object co-localization technique first discussed in~\cite{Duchenne2011,Duchenne2012} and takes as input pairs of flattened cortical surfaces from participants who previously took part in an arbitrary fMRI experiment that exposed them to complex, varying stimuli (e.g. visual categorization~\cite{Iordan2015}). We standardized the cortical surfaces by resampling the multidimensional functional data of each subject to a regular square grid at a resolution of 2mm x 2mm. Each point in the resulting grids has a functional time course associated with it which corresponds to the estimated response of that point to the stimuli shown across the entire duration of the fMRI experiment (e.g. using a 512 TR fMRI experiment as input implies a 512-dimensional representation for each point in the resulting standardized cortical maps). Then, for each possible pair of participants, one of them is selected as the seed and the other as the target (for our final results, each participant in each pair is, in turn, selected as the seed and target, and performance is averaged across both these configurations). The location of the functionally defined region of interest in the seed subject is then tiled with a grid of $n$ x $n$ patches, where each patch is associated with a small area on the brain surface (e.g. 5 x 5 voxels). Finally, the algorithm seeks to find maximal functional correspondences between each seed patch and an equivalent region in the target map by maximizing the sum of time course correlations across all the patches, while enforcing that the distances between adjacent patches change by less than a specified amount in each direction (i.e. $\rho = 4$ voxels) between the seed and target maps. An example seed ROI parcellation and target matching are shown in Fig.~1 (right). The optimization problem can be written as:

\vspace{-5mm}
\begin{eqnarray*}
\underset{M}{\minimize} &  \sum_{i} d_F (F_i, F_{m_i}) \\
\st & d_s (p_{m_i}, p_{m_j}) \leq \rho,
\end{eqnarray*}

\noindent where $M = \{ (i, m_i) \}$ is the collection of correspondences between seed ($i$) and target patches ($m_i$); $d_F$ is the feature distance between the patches in each correspondence, computed as 1 - Pearson $r$; $d_s$ is the cortical distance difference between the original and mapped configuration of each pair of patches (patch $i$ mapped to patch $j$) in the two maps; and $\rho$ is the maximum allowable distance change between neighboring patches across maps. We solve the optimization problem above using a deterministic grid search through the space of all possible patch jitter permutations.

\subsection{Advantages Over Previous Methods}

Our method presents several key advantages over other alignment methods, which render it more general and more precise. First, virtually all previous methods compute a complete correspondence code between entire cortical surfaces. Afterwards, the location of functional areas is obtained second-hand, e.g. by aligning a contrast map and re-thresholding. Here, we instead focus on maximizing the quality of the mapping for a single, specific seed ROI. Furthermore, other alignment methods usually generate a smooth manifold transformation between cortices. However, this entails a very strong assumption that activation profiles vary smoothly and with the same spatial distribution across subjects. We forgo this assumption by allowing locally-non-smooth deformations in the topology of the predicted ROI. Finally, cortical registration methods are usually described by highly complex optimization problems that can only be solved up to a local minimum, and are thus highly sensitive to parameter initialization. By contrast, our method has a global optimum solution to which we converge deterministically and is therefore much more robust.

\section{Experiments}

\subsection{fMRI Dataset and Baselines}

We tested our method by predicting the location of a difficult to match, functionally defined, object-selective ROI (lateral occipital complex LOC) between subjects using data from a block design passive-viewing fMRI experiment where participants ($n=7$) were shown 1,024 images of objects from 32 categories (Fig.~2, see~\cite{Iordan2015} for details about the procedure and preprocessing). We computed the position of each participant's LOC using standard localizer runs conducted in a separate fMRI session~\cite{Golarai2007,Sayres2008}. We then used the AFNI-SUMA software package~\cite{AFNI} to project and interpolate the data from the 3D volume onto a 2D flattened regular grid cortical map.

We compared our algorithm against the two most commonly used cortical registration methods: anatomical landmark-based AFNI 3dvolreg~\cite{AFNI} and cortical convexity-based FreeSurfer~\cite{FreeSurfer}. AFNI uses information about overall brain shape and automatically defined anatomical points of interest to warp cortical volumes across subjects. FreeSurfer also uses brain shape, as well as information about cortical curvature (sulci and gyri locations, distribution of normals to the gray matter surface) to iteratively distort one cortical surface into another.
 
\begin{figure}[t]
\begin{center}
\includegraphics[width=0.75\textwidth]{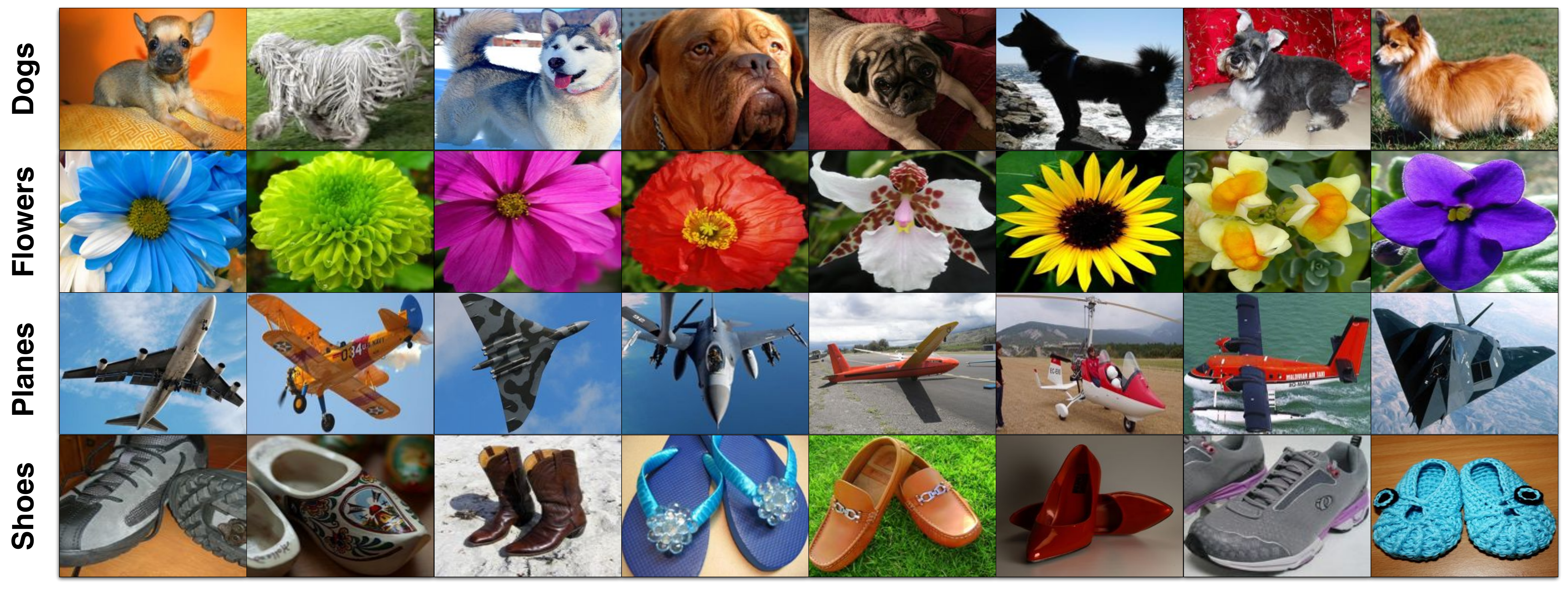}
\caption{\label{stimuli} Stimulus set for fMRI experiment used to perform and evaluate the cortical prediction algorithm. During the experiment, participants were shown images from 32 object categories: {8 breeds of dogs, 8 types of flowers, 8 types of planes, 8 types of shoes} (32 images per category; 1,024 images total).}
\end{center}
\end{figure}

\begin{figure*}[t]
\begin{center}
   \includegraphics[width=0.9\textwidth]{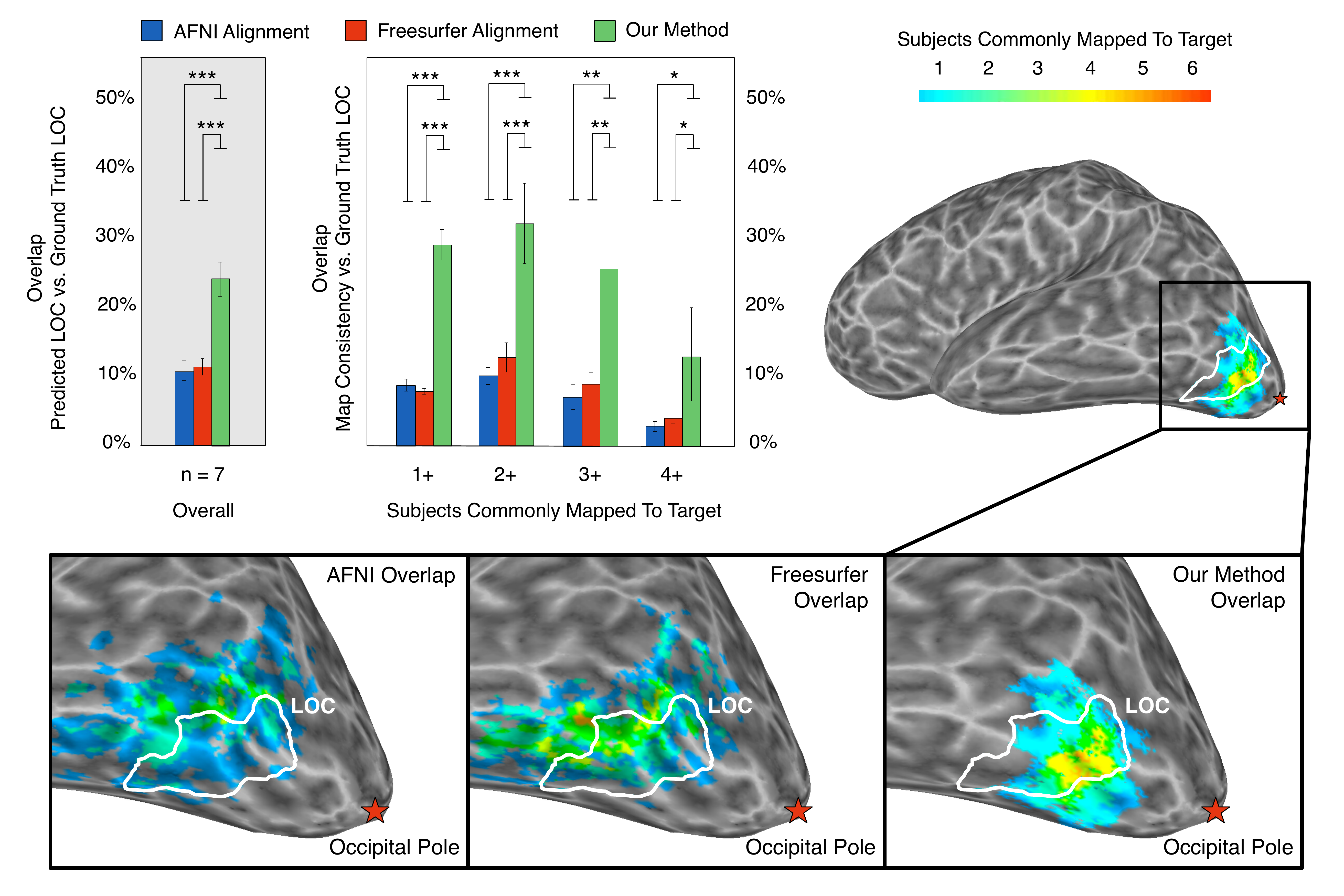}
\end{center}
   \caption{{\bf Alignment Results: Accuracy and Consistency} ($n=7$ subjects). For every target subject, we align LOC from all other 6 subjects to the target cortical surface using functional data from the above experiment. {\bf (Top Left)} Overlap between predicted LOC and LOC defined using separate standard localizer procedure, measured as intersection over union of surfaces. {\bf (Top Right)} We select the voxels predicted consistently in the target map for $n+$ subjects and compute the overlap between this restricted region and ground truth LOC for $n \in \{ 1, 2, 3, 4 \}$. {\bf (Bottom)} Consistency of predicted LOC obtained from aligning using AFNI 3dvolreg, FreeSurfer, and Our Method for a representative subject. Heatmap indicates how many subjects' LOC were mapped to that voxel on the target surface. White outline indicates LOC boundaries defined using separate standard localizer procedure.}
\label{fig:supp}
\end{figure*}

\subsection{Results}

To test how well our method predicts the location of LOC across subjects, we used two metrics: accuracy and consistency. Accuracy represents the percentage of overlap between functionally-defined LOC and predicted LOC after mapping from a different subject's brain. Consistency is defined as the amount of overlap between predicted regions from multiple subjects aligned to the same target map. For both metrics, overlap is computed as intersection over union.

We show results for the two baselines, as well as our method in Fig.~\ref{fig:supp}. Our registration method vastly outperformed the two canonical baselines in overlap between predicted region and ground truth LOC: baselines 10-11\%, ours 24-25\%. Furthermore, the maps obtained using our method are more consistent across subjects than both baseline measures (overlap of region commonly mapped from 3+ subjects: baselines 9-11\%, ours 26\%).

Qualitatively, the cortical maps further showcase the strength of our results compared to the AFNI and FreeSurfer baselines. In the first two panels of Fig.~\ref{fig:supp} (bottom left) we see that functional regions in other subjects are mapped with a high degree of variance onto the target subject cortical sheet. Often, there is little overlap with our localizer-defined ROI and, most importantly, the mapping may place the region several centimeters away from its desired location, often on a different gyrus. By contrast, our method (Fig.~\ref{fig:supp}, bottom right) shows much less variance in the predicted area, with the peak of the prediction fully contained within our localizer-defined region.

These results suggest that our registration technique significantly increases the reliability of transferring the location of functional ROIs between subjects.

\section{Conclusion}

In this paper, we proposed a locally optimized registration method that predicts the location of a seed region of interest (ROI) on a separate target cortical sheet by maximizing the functional correlation between regions and simultaneously constraining the global structure of the mapping, while allowing for non-local deformations in its topology.

Our method vastly outperforms two canonical alignment baselines (anatomical landmark based AFNI \cite{AFNI} and cortical curvature based FreeSurfer~\cite{FreeSurfer}) in both precision and consistency. By improving the quality and reliability of matching and transferring the location of functional ROIs across subjects, our technique represents an important step towards obviating the need for running separate time- and resource-consuming localizer scans for every functional brain region. Instead, we envision an eventual solution where a single 'localizer' experiment is performed using a high variance stimulus (i.e. natural movie~\cite{Huth2012}), which is then used to define all functional ROIs, including potential regions which have yet to be identified. Such a mapping is also useful in settings where one needs to compare analyses and hypotheses between datasets where functional localizers are missing and gathering extra sessions of data is either expensive (large number of participants) or impossible (unavailability of former subjects).

Finally, the relationship between peaks of functional contrasts and the computation performed by the cortex surrounding them is not well understood. Since our method improves the quality of functional ROI mapping between subjects, it becomes especially useful for investigating the key complex relationship between anatomy, functional contrast peaks, and cortical computation.

\bibliographystyle{ieeetr}
{\bibliography{Iordan_etal_MLINI_2015_bib}}

\end{document}